\documentstyle[aps]{revtex}

\begin{document}
\draft
\title{Statistical mechanics and path integrals for a finite number of bosons.}
\author{L. F. Lemmens}
\address{Departement Natuurkunde, Universiteit Antwerpen (RUCA),\\
Groenenborgerlaan 171, B-2020 Antwerpen}
\author{F.~Brosens 
\thanks{Senior Research Associate of the FWO (Fonds voor Wetenschappelijk Onderzoek-Vlaanderen).}%
and J.T. Devreese 
\thanks{Also at Universiteit Antwerpen (RUCA) and Technische Universiteit Eindhoven, Nederland.}%
}
\address{Departement Natuurkunde, Universiteit Antwerpen (UIA),\\
Universiteitsplein 1, B-2610 Antwerpen}
\date{November 13, 1997}
\maketitle

\begin{abstract}
Recent investigations show that the statistical mechanics of a finite number
of particles in ideal harmonic systems predicts different results for the
same physical properties, depending on the ensemble under consideration.
Path integral methods for a finite number of bosons with equidistant energy
levels give the same answers for the mean energy, the specific heat and the
condensation temperature etc., irrespective whether their calculation
results from the density of states, from the partition function or from the
generating function.

We show that this contradiction is due either to the use of approximate
relations between quantum statistical expressions, or to a misinterpretation
of the generating function.
\end{abstract}


Since Bose--condensed vapors became available to experimentation \cite
{AndSc95,DavPRL95,BraPRL95,BraPRL97}, considerable theoretical interest has
been raised in the statistical mechanics of an idealization of the vapor,
i.e. a system with equally spaced energy levels for a finite number of
particles. Conventional statistical mechanics for a system in equilibrium at
a given temperature (expressed by $\beta =1/kT$ ) states that the
probability of the system having energy $E_n$ is proportional to $\exp
\left( -\beta E_n\right) .$ The calculation of the energies $E_n$ is a
quantum mechanical problem. Equally spaced energy levels are justified by
the parabolic confinement potential. The quantum statistical theory takes
into account the discreteness of the levels and the limited number of
particles and contrasts therefore with earlier studies of Bose--Einstein
condensation, because the thermodynamic limit and the quasi--continuity of
the levels cannot be used as a justification \cite{Huang}. In the path
integral approach to quantum statistical theory \cite{FeyBen72,Schulman},
the calculation of the levels as well as the density of states or the
partition function form an intrinsic part of the same study and are obtained
simultaneously, in contradistinction with the approach where a quantum
problem gives the energy levels and where subsequently statistical theory is
used to study the cooperative behavior.

For identical particles the conditional probability to find a system
containing $N$ particles in the neighborhood of the configuration $r$ can be
calculated by the path integral method \cite{BDLPRE97a,BDLPRE97b}. We have
calculated the partition function $Z\left( \beta ,N\right) $ and some static
response functions of a Gaussian model for bosons as well as for fermions
using this approach. The related thermodynamical quantities have been
obtained from the free energy that was calculated using the probability
generating function $\Xi \left( \beta ,\gamma \right) $. The partition
function $Z\left( \beta ,N\right) $ relates to the density of states $\Omega
\left( E,N\right) $ as follows: 
\begin{equation}
Z\left( \beta ,N\right) =\int e^{-\beta E}\Omega \left( E,N\right) dE,
\end{equation}
while the generating function is defined by: 
\begin{equation}
\Xi \left( \beta ,\gamma \right) =\sum_N\gamma ^NZ\left( \beta ,N\right) .
\label{Ksigen}
\end{equation}
It should be remarked that the real variables $\gamma $ and $\beta $ used to
interrelate the three quantities $\Omega \left( E,N\right) ,$ $Z\left( \beta
,N\right) $ and $\Xi \left( \beta ,\gamma \right) $ are independent
variables \cite{Kubo}. The inversion formula to obtain $\Omega \left(
E,N\right) $ from $Z\left( \beta ,N\right) $ is the inverse Laplace
transform 
\begin{equation}
\Omega \left( E,N\right) =\frac 1{2\pi i}\int\limits_{\beta -i\infty
}^{\beta +i\infty }e^{\tau E}Z\left( \tau ,N\right) d\tau ,
\end{equation}
where $\beta $ is larger than the real part of all poles of $Z\left( \tau
,N\right) $ in the complex $\tau $ plane. For the inversion of $\Xi \left(
\beta ,\gamma \right) $ the residue theorem can be used: 
\begin{equation}
Z\left( \beta ,N\right) =\frac 1{2\pi i}\oint \frac{\Xi \left( \beta
,z\right) }{z^{N+1}}dz.  \label{eq:Zcontour}
\end{equation}

It is generally accepted that the probabilities associated with events by
statistical mechanics, are consistent with those of quantum statistical
theory \cite{ZipPR77}. Referring to the normalization factor $Q=\sum_{n}\exp
\left( -\beta E_{n}\right) $ of the probability in statistical mechanics as
the canonical partition function, it is implicitly assumed that $Q$ and $Z$
are the same for the same system, and lead consequently to the same
predictions. Therefore it should not matter which one of the three
quantities $\Omega ,\ Z$ or $\Xi $ is calculated because of their exact
interrelationship. {\em Quantum statistical theory would be in conflict with
statistical mechanics} if in statistical mechanics the quantities
corresponding with $\Omega ,\ Z$ and $\Xi $ (usually indicated by their
ensemble: microcanonical, canonical or grand canonical), would lead to
different predictions, what they do according to \cite{many}. Recently this
puzzle even gave rise to the introduction of a new {\em fourth} ensemble,
coined the ``Maxwell Demon'' ensemble \cite{GajPRL97,NavPRL97}, with the aim
to take the ensemble conditions for condensation better into account. The
quantum statistical analogue of this ``Maxwell Demon'' ensemble is obtained
by performing the inversions starting from $\Xi $ in the appropriate order 
\begin{equation}
\Upsilon \left( E,\gamma \right) =\frac{1}{2\pi i}\int\limits_{\beta
-i\infty }^{\beta +i\infty }e^{\tau E}\Xi \left( \tau ,\gamma \right) d\tau
=\sum_{N}\Omega \left( E,N\right) \gamma ^{N},
\end{equation}
expressing also this quantity in a unique way in terms of the density of
states.

The main question that will be addressed in the present letter is: {\em what
is the origin of the apparent discrepancies between predictions based on
statistical mechanics }? We will discuss here two possibilities. The first
one is based on probability considerations. The second one elaborates upon
an old warning of Zip, Uhlenbeck and Kac \cite{ZipPR77} against grand
canonical partition functions calculated from $\Xi \left( \beta ,e^{\beta
\mu }\right) =Tr\left[ \exp \left( -\beta \left( H-\mu N\right) \right)
\right] .$

\subsubsection{A conditional probability density approach.}

First we show how the discrepancies mentioned above are induced by an
approximation which is justified in the limit of an arbitrarily large number
of particles, but not for a finite number of particles. We do this on the
basis of our approach \cite{BDLPRE97a} that we briefly describe here.
Denoting by $r$ the configuration $\left( {\bf r}_1,{\bf r}_2,\ldots ,{\bf r}%
_N\right) $ of the $N$ particles, the potential energy function $V\left(
r\right) $ of our model is given by 
\begin{equation}
V\left( r\right) =\frac m2\Omega ^2\sum_{j=1}^N{\bf r}_j^2-\frac{m\omega ^2}2%
\sum_{j<k}^N\left( {\bf r}_j-{\bf r}_k\right) ^2.  \label{eq:potential}
\end{equation}
Because the model is Gaussian the propagator for the particles (considered
distinguishable) is known. Using path integral methods the projection on the
symmetric boson states (antisymmetric for fermions) could be performed,
leading to an expression for the partition function $Z_B\left( \beta
,N\right) $ for bosons. The center-of-mass contribution factorizes out,
leaving $3\left( N-1\right) $ internal degrees of freedom with frequency $w=%
\sqrt{\Omega ^2-N\omega ^2}$. Introducing $b=e^{-\beta \hbar w}$, the
generating function {\em for these internal degrees of freedom} can be
written as 
\begin{equation}
\Xi _B\left( \gamma ,\beta \right) =\exp \left[ \sum_{\ell =1}^\infty \frac{%
\gamma ^\ell }\ell \left( \frac{b^{\frac 12\ell }}{1-b^\ell }\right)
^3\right] ,  \label{eq:Ksiexplicit}
\end{equation}
and is the result of a functional integration. Another but more familiar
representation of the same function is: 
\begin{equation}
\Xi _B\left( \gamma ,\beta \right) =\prod_{\nu =0}^\infty \left( \frac 1{%
1-\gamma b^{{\frac 32}+\nu }}\right) ^{\frac 12(\nu +1)(\nu +2)}.
\end{equation}
The same function can be obtained directly using combinatorial analysis
within the assumptions of statistical mechanics.\cite
{Grossman,Kirsten,Ketterle,Grossman2,Haugerud,vanDru97}. In our case $\Xi
_B\left( \gamma ,\beta \right) $ is only formally the grand canonical
partition function of a set of identical particles in a parabolic well. To
become really a grand canonical partition function the substitution $\gamma
\rightarrow \exp (\mu \beta )$ has to be made and a calculation procedure
for $\mu $ has to be given. If $N$ is sufficiently large, stationary phase
or steepest descent methods can be used to {\sl approximate} the inversion
formulas \cite{Kubo} and give a meaning to the chemical potential $\mu $.

As an illustration, let us consider the partition function $Z_B\left( \beta
,N\right) $ based on the contour integral (\ref{eq:Zcontour}) for the
specific example of harmonically interacting bosons in an harmonic
confinement potential, i.e. the actual calculations are performed using the
explicit form (\ref{eq:Ksiexplicit}). The generating function $\Xi _B\left(
z,\beta \right) $ makes the direct numerical evaluation of (\ref{eq:Zcontour}%
) unfeasible. The interested reader is referred to \cite{BDLPRE97a,BDLPRE97b}
for a discussion. However, considering a circular contour around the origin
with radius $u,$ the substitution $z=ue^{i\theta }$ transforms the contour
integral into 
\begin{equation}
Z_B\left( \beta ,N\right) =\frac 1{2\pi }\int_0^{2\pi }\frac{\Xi \left(
ue^{i\theta },\beta \right) }{u^N}e^{-iN\theta }d\theta =\frac 1{2\pi }%
\int_0^{2\pi }e^{\left[ \ln \Xi \left( ue^{i\theta },\beta \right) -N\ln
u\right] }e^{-iN\theta }d\theta .
\end{equation}
The extremum of $\left[ \ln \Xi \left( ue^{i\theta },\beta \right) -N\ln
u\right] $ on the real axis satisfies the condition 
\begin{equation}
N=u\frac d{du}\ln \Xi \left( u,\beta \right) ,
\end{equation}
which is precisely the expression for the expected number of particles $N$
in the ``grand canonical ensemble'' if $u$ is interpreted as $u=e^{\beta \mu
}$. Factorizing out this steepest descent contribution and extracting the
real part, one obtains with little effort 
\begin{eqnarray}
Z_B\left( \beta ,N\right) &=&Z_B^{\left( 0\right) }\left( \beta ,N\right)
\int_0^\pi \Psi \left( \theta \right) d\theta , \\
Z_B^{\left( 0\right) }\left( \beta ,N\right) &=&\frac{\Xi _B\left( u,\beta
\right) }{u^N}, \\
\Psi \left( \theta \right) &=&\frac 1\pi \frac{\Xi _B\left( ue^{i\theta
},\beta \right) }{\Xi _B\left( u,\beta \right) }e^{-iN\theta }.
\label{eq:Psi}
\end{eqnarray}
The corresponding free energy thus becomes the sum of two contributions 
\begin{eqnarray}
F_B\left( \beta ,N\right) &=&F_B^{\left( 0\right) }\left( \beta ,N\right) -%
\frac 1\beta \ln \left( \int_0^\pi \Psi \left( \theta \right) d\theta
\right) , \\
F_B^{\left( 0\right) }\left( \beta ,N\right) &=&-\frac 1\beta \ln \frac{\Xi
_B\left( u,\beta \right) }{u^N}.
\end{eqnarray}
$F_B^{\left( 0\right) }\left( \beta ,N\right) $ is the result which one
would obtain in the ``grand canonical treatment''. The remainder $\frac 1%
\beta \ln \left( \int_0^\pi \Psi \left( \theta \right) d\theta \right) $ can
be obtained by integration and this correction is crucial for a finite
number of particles. The integrand $\Psi \left( \theta \right) $ is shown in
Fig. 1 for $N=10$ as a function of $\theta $ for various temperatures $T$,
expressed in units of the condensation temperature $T_c=\left( N/\zeta
\left( 3\right) \right) ^{1/3}\hbar w/k$. Below the condensation temperature
the oscillations in $\Psi \left( \theta \right) $ are distributed more or
less uniformly (for bosons) over the integration interval $\left[ 0,\pi
\right] $ and above the condensation temperature they are strongly damped.
The difference between $F_B\left( \beta ,N\right) $ and the zero-order
approximation $F_B^{\left( 0\right) }\left( \beta ,N\right) $ decreases with
an increasing number of particles. The results for the free energy are shown
in Fig. 2--4 for $N=1,$ $10$ and $100,$ where $f\left( \beta ,N\right)
\equiv F_B\left( \beta ,N\right) /N\hbar w$ is plotted versus $T/T_c.$ For
comparison, the zero-order approximation $F_B^{\left( 0\right) }\left( \beta
,N\right) /N\hbar w$ is also plotted. The numerical results for $F_B\left(
\beta ,N\right) $ are to within 6 digits in agreement with those obtained
earlier \cite{BDLPRE97a} from a recurrence relation for the partition
function.

If the partition function obtained using our path integral method relates to
the {\em conditional probability function} for the energy density {\em given}
the number of particles, we may summarize that the predictions based on the
canonical partition function in statistical mechanics are {\em not in
conflict} with quantum statistical theory, provided the density of states
(microcanonical) is correctly derived by the inversion formulas. This
concludes our discussion based on the conditional probability interpretation
of the density of states.

\subsubsection{The joint probability density approach}

It is a common error to interpret a conditional probability as a {\em joint
probability function. } In order to analyse the consequences of this
possibility, let us {\em \ assume} that the function $\Xi _B\left( \gamma
,\beta \right) $ is the generator of such a joint probability function for
the energy $E$ {\em and} the number particles $N$, i.e. we give ourself a
marginal distribution for the number of particles consistent with that
generating function. The theory of continuous-time Markov processes for a
queue with infinitely many servers\cite{BhatWay} might illustrate this
argument. In this stochastic model of a facility, customers are entering
according to a Poisson distribution and are subsequently served with an
exponentially distributed service time. In the case that both distributions
(of the customers and of the servers) have the same parameter, the
probability $p_N(t)$ that $N$ customers are waiting to be served at time $t$
can be calculated from the generator ${\cal A}$ of the process. This
generator explicitly takes the transitions to one more or to one less
customer into account. Denoting $p_N(t)=\left\langle N\left| e^{t{\cal A}%
}\right| N\right\rangle $ and $g\left( z,t\right) =\sum_{N=0}^\infty
z^Np_N(t),$ the analogy between the probability generating function of the
stochastic process and the generating function for the partition function
becomes obvious.

We will discuss the particle fluctuations of the ``grand canonical
ensemble'' in this interpretation and elucidate the remark of ref.\cite
{ZipPR77} earlier cited on that ensemble. In order to do so it is
instructive to introduce an alternative to invert a power series. This
inversion will allow to make a connection between the generating function in
this interpretation and coherent states. Using 
\begin{equation}
\frac 1{N!}\frac 1\pi \int e^{-\left| z\right| ^2}\bar{z}^Nz^Ld^2z=\delta
_{N,L}\,,  \label{eq:orthogonal}
\end{equation}
where $\bar{z}$ denotes the complex conjugate of $z,$ one obtains the
following expression for the partition function $Z\left( \beta ,N\right) $
from $\Xi \left( z,\beta \right) $: 
\begin{equation}
Z\left( \beta ,N\right) =\frac 1{N!}\frac 1\pi \int e^{-\left| z\right| ^2}%
\bar{z}^N\Xi \left( z,\beta \right) d^2z.
\end{equation}
Filling out the power series (\ref{Ksigen}) and summing over $N$ one finds: 
\begin{equation}
\sum_{N=0}^\infty Z\left( \beta ,N\right) =\frac 1\pi \int \sum_{N=0}^\infty
\sum_{L=0}^\infty \frac{\bar{z}^Ne^{-\frac 12\left| z\right| ^2}}{\sqrt{N!}}%
Z\left( \beta ,L\right) \frac{z^Le^{-\frac 12\left| z\right| ^2}}{\sqrt{L!}}%
d^2z,
\end{equation}
where use has been made of the orthogonality relation (\ref{eq:orthogonal})
to replace a denominator $N!$ by $\sqrt{N!L!}.$

Introducing formally a Hamiltonian ${\cal H}$ that allows for transitions
between systems with different numbers of particles, the following form 
\begin{equation}
\sum_{N=0}^\infty Z\left( \beta ,N\right) =\frac 1\pi \int \left\langle
z\left| e^{-\beta {\cal H}}\right| z\right\rangle d^2z
\end{equation}
helps to recognize the coherent state representation for the normalization
of the probability for the event that ``the system contains $N$ particles
given the inverse temperature $\beta $''. The state $\left| z\right\rangle $
can be considered as a coherent matter state built up from Bose systems
containing a different number of particles: 
\begin{equation}
\left| z\right\rangle =\sum_N\frac{z^N}{\sqrt{N!}}e^{-\frac 12\left|
z\right| ^2}\left| N\right\rangle .
\end{equation}
The derivation of the actual form of the Hamiltonian ${\cal H}$ needed to
study other than equilibrium properties is beyond the scope of this letter.
In order that the equilibrium properties predicted by ${\cal H}$ coincide
with those derived from the generating function, ${\cal H}$ should satisfy
the condition $\sum_L\left\langle N\left| {\cal H}\right| L\right\rangle 
\sqrt{\frac{L!}{N!}}=0,$ which expresses the conservation of probability 
\cite{VKam92,Jau90}, for a generator of a stochastic process. When the
equilibrium properties of a model with fluctuations in the number of
particles are considered, information on this model can be extracted from
the generating function $\Xi \left( z,\beta \right) $ by making the
following identification 
\begin{equation}
\left\langle z\left| e^{-\beta {\cal H}}\right| z\right\rangle =e^{-\bar{z}%
\left( z-1\right) }\Xi \left( z,\beta \right) .  \label{idcoh}
\end{equation}
The analogy with the queue suggests that if the generating function derives
from a density of states that is interpreted as a joint probability density,
it leads to a marginal distribution of the fluctuating number of particles.
Also the assumed stochastic behavior requires a specific form for the
transitions between a well with $N$ bosons and one with $N\pm 1$ bosons.
These ``a priori'' properties of that model are certainly not satisfied in
the experimental set-up \cite{AndSc95,DavPRL95,BraPRL95,SilWal81} where the
number of particles are monitored. Therefore, interpreting $\Xi \left(
z,\beta \right) $ given by (\ref{eq:Ksiexplicit}) as the generator of a
joint probability density does in our opinion not allow to draw conclusions
on the Bose condensed systems, discussed in the introduction. Changing the
order of inversion to obtain the ``Maxwell demon'' distribution without
changing the interpretation will not alter our objections. This means that
the study of a Bose condensed system with respect to a fluctuating number of
particles requires a quantum statistical theory incorporating explicitly the
transitions between states containing a different number of bosons, what is
not the case in the quantum statistical theory leading to (\ref
{eq:Ksiexplicit}).

\subsubsection{Conclusion}

In general we conclude that for a given number of particles the concepts put
forward in \cite{FeyBen72} and worked out partially in \cite{BDLPRE97a} and 
\cite{BDLPRE97b} for systems of identical particles with equally spaced
energy levels, allow to trace back differences linked with the statistical
ensembles to different {\sl approximations} in the inversion of the
generating function. These differences have been quantified for the free
energy and are identified with oscillatory behavior in our integrand $\Psi
\left( \theta \right) $. At least for indistinguishable (identical)
particles, these oscillations become negligible in the large $N$ limit. It
should be stressed that the large $N$ behavior is a direct consequence of
the projection on the symmetric (antisymmetric) representation of the
permutation group ensuring indistinguishability of the identical particles.
Without the projection on the symmetric irreducible representation, this
large $N$ behavior for the oscillations is absent. In that case the relative
error made by replacing the integral by its steepest descent approximation 
{\em does not} become negligible. This observation raises a new challenging
question in the statistical description of identical particles: is the
projection on the symmetric (antisymmetric) representation necessary in all
circumstances?

\acknowledgments

Discussions with W. Krauth, F. Lalo\"{e} and Y. Kagan are acknowledged. This
work is performed in the framework of the FWO projects No. 1.5.729.94, G.
0287.95 and WO.073.94N (Wetenschappelijke Onderzoeksgemeenschap, Scientific
Research Community of the FWO on ``Low-Dimensional Systems''), the
``Interuniversity Poles of Attraction Program -- Belgian State, Prime
Minister's Office -- Federal Office for Scientific, Technical and Cultural
Affairs'', and in the framework of the BOF NOI 1997 projects of the
University of Antwerpen. Part of this work (L.F.L.) has been performed under
the RAFO--contract 61219.

\newpage {\bf Figure captions}

\begin{description}
\item[Fig. 1:]  Integrand $\Psi \left( \theta \right) $ [see (\ref{eq:Psi})]
for the remainder term of the partition function, calculated for $10$ bosons
at various temperatures.

\item[Fig. 2:]  Scaled free energy $f\left( \beta ,N\right) \equiv F_B\left(
\beta ,N\right) /N\hbar w$ as a function of $T/T_c$ for $N=1$. For
comparison, the zero-order steepest descent approximation $F_B^{\left(
0\right) }\left( \beta ,N\right) /N\hbar w$ is also plotted (dashed line).

\item[Fig. 3:]  Same as Fig. 2, for $N=10$ bosons.

\item[Fig. 4:]  Same as Fig. 3, for $N=100$ bosons.
\end{description}


\begin{references}
\bibitem{AndSc95}  M. H. Anderson, J. R. Ensher, M. R. Matthews, C. E.
Wieman, and E. A. Cornell, {\sl Science }{\bf 269, }198 (1995).

\bibitem{DavPRL95}  K. B. Davis, M. O. Mewes, M. R. Andrews, N. J. van
Druten, D. S. Durfee, D. M. Kurn, and W. Ketterle, Phys. Rev. Lett. {\bf 75}%
, 3969 (1995).

\bibitem{BraPRL95}  C. C. Bradley, C. A. Sacket, J. J. Tollett, and R. G.
Hulet, Phys. Rev. Lett. {\bf 75}, 1687 (1995).

\bibitem{BraPRL97}  C. C. Bradley, C. A. Sacket, and R.G. Hulet, {\sl \ }%
Phys. Rev. Lett. {\bf 78}, 985 (1997).

\bibitem{Huang}  K. Huang, {\sl Statistical Mechanics} (J. Wiley, New York
,1987).

\bibitem{FeyBen72}  R.P. Feynman, {\sl Statistical Mechanics, a Set of
Lectures} (W$.$ A. Benjamin Inc., Reading, 1972).

\bibitem{Schulman}  L. Schulman {\sl Techniques and Applications of Path
Integration }(J. Wiley and Sons, New York, 1981)

\bibitem{BDLPRE97a}  F. Brosens, J. T. Devreese, and L. F. Lemmens, Phys.
Rev. E{\bf \ 55}, 227 (1997).

\bibitem{BDLPRE97b}  F. Brosens, J. T. Devreese, and L. F. Lemmens, Phys.
Rev. E {\bf 55}, 6795{\bf \ }(1997).

\bibitem{Kubo}  R. Kubo, {\sl Statistical Mechanics} (North Holland,
Amsterdam, 1965).

\bibitem{ZipPR77}  R.M. Zipp, G.H. Uhlenbeck, and M. Kac, Phys. Rep. {\bf %
\thinspace 32}, 169 (1972).

\bibitem{many}  S. Grossmann and M. Holthaus, Phys. Lett. A {\bf 208},188
(1995); \newline
S. Grossmann and M. Holthaus, Phys. Rev. E{\bf \ 54}, 3495 (1996); \newline
H. D. Politzer, Phys. Rev. A {\bf 54}, 5048{\bf \ }(1996); \newline
C. Herzog and M. Olshanii, Phys. Rev. A{\bf \ 55}, 3254 (1997).

\bibitem{GajPRL97}  M. Gajda, K. Rzazewski, Phys. Rev. Lett. {\bf 78}, 2686
(1997).

\bibitem{NavPRL97}  P. Navez, D. Bitouk, M. Gajda, Z. Idraszek, and K.
Rzazewski, Phys. Rev. Lett. {\bf 79}, 1789 (1997).

\bibitem{Grossman}  S. Grossman and M. Holthaus, Z. Naturforsch. {\bf 50a},
323; 921 (1995).

\bibitem{Kirsten}  K. K. Kirsten and D. J. Toms, Phys. Rev. A {\bf 54}, 4188
(1996).

\bibitem{Ketterle}  W. Ketterle and N. J. van Druten, Phys. Rev. A {\bf 54},
656 (1996).

\bibitem{Grossman2}  S. Grossman and M. Holthaus, Phys. Rev. Lett. {\bf 79},
3557 (1997)

\bibitem{Haugerud}  H. Haugerud, T. Haugset and F. Ravndal, Phys. Lett. A. 
{\bf 225}, 18 (1997).

\bibitem{vanDru97}  N. J. van Druten and W. Ketterle, Phys. Rev. Lett. {\bf %
79}, 549 (1997).

\bibitem{BhatWay}  R. N. Bhattacharya and E. C. Waymire, {\sl Stochastic
Processes with Applications} (J. Wiley \& Sons, New York, 1990).

\bibitem{VKam92}  N. G. Van Kampen, {\sl Stochastic Processes in Physics and
Chemistry} (North Holland, Amsterdam, 1992).

\bibitem{Jau90}  H. R. Jauslin, Phys. Rev. A {\bf 41}, 3407 (1990).

\bibitem{SilWal81}  I. F. Silvera and J. T. M. Walraven, J. Appl. Phys. {\bf %
52}, 2304 (1981).
\end{references}
\end{document}